\documentclass[twocolumn, prl, showpacs]{revtex4}
\usepackage{epsfig}
\graphicspath{{Fig/}}
\usepackage{amsmath}
\usepackage{amsfonts}
\usepackage{amssymb}
\usepackage{graphicx}
\usepackage{subfigure}

\begin{document}

\title{Temperature dependence of  normal mode reconstructions of protein dynamics}

\author{Francesco Piazza} 
\author{Paolo De Los Rios}
\affiliation{Laboratoire de Biophysique Statistique, SB ITP ,\\ 
             Ecole Polytechnique F\'ed\'erale de Lausanne - EPFL, CH-1015, Lausanne, Switzerland}
\author{Fabio Cecconi}
\affiliation{SMC-INFM Center for Statistical Mechanics and Complexity (CNR) 
             and Istituto dei Sistemi Complessi CNR\\
             Via dei Taurini 19, 00185 Rome, Italy.}
\date{\today} 

\begin{abstract}
Normal mode analysis is a widely used technique for reconstructing 
conformational changes of proteins from the knowledge of native structures. 
In this Letter, we investigate to what extent normal modes capture
the salient features of the dynamics over a range of temperatures from 
close to $T=0$ to above unfolding. 
We show that  
on the one hand, the use of normal modes at physiological temperatures 
is justified provided proteins are cooperative. On the other hand, 
it is imperative to consider several modes in order to eliminate the 
unpredictable temperature dependence of single-mode contributions to the 
protein fluctuations.
\end{abstract} 

\pacs{87.15.-v, 87.10.Tf, 87.14.E-}

\maketitle

The effectiveness of proteins at performing their functions strongly 
depends on their ability to dynamically explore different conformations. 
In many cases, the different functionally relevant conformations 
correspond to global deformations 
of the molecules, although not necessarily of large amplitude.
In such circumstances, the small-amplitude dynamics
can be adequately described through  Normal Modes (NM) analysis
in order to decipher the  structure-dynamics-function relation~\cite{Bahar2005a,Case1994}.


NM analysis of macromolecules
has a venerable history, dating back to the early 
1980's~\cite{Go1983, Brooks1983a}, and has
enjoyed several renaissances, most prominently in its latest variant, the
Anisotropic Network Model and its derivatives \cite{Tirion1996,Atilgan2001}. 
The cornerstone of NM approach is the
assumption that functional dynamics of proteins can be captured  
by the lowest-frequency harmonic normal 
modes as long as the native-state deformations are of small amplitude. 
Therefore, the complex potential energy $V(\lbrace \vec{r} \rbrace)$
ruling protein dynamics can be expanded, 
as a function of the atomic coordinates $\vec{r}_i$, $i=1,\dots,N$,  
to second order in the fluctuations around the native state
\begin{eqnarray}
V(\lbrace \vec{r} \rbrace) &\simeq& \frac{1}{2} 
\sum_{i,j} \partial^2_{i\alpha,j\beta} V |_{\vec{R}_{i}}  
\Delta r_{i\alpha} \Delta r_{j\beta}\;,
\label{eq:harmonic}   
\end{eqnarray}
where $\Delta r_{i\alpha} = r_{i\alpha}-R_{i\alpha}$, 
$R_{i\alpha}$ being the native state coordinates. The dynamical 
matrix of the system is defined as $D = M^{-1/2} H M^{-1/2}$, with $M$ the
diagonal mass matrix and $H$ the Hessian of the potential. 
The eigenvectors $\hat{\psi}_k$, $k=1,...,3N$, of $D$ are the normal modes of the protein.
\begin{figure}[b!] 
\centering
\includegraphics[width=\columnwidth,clip]{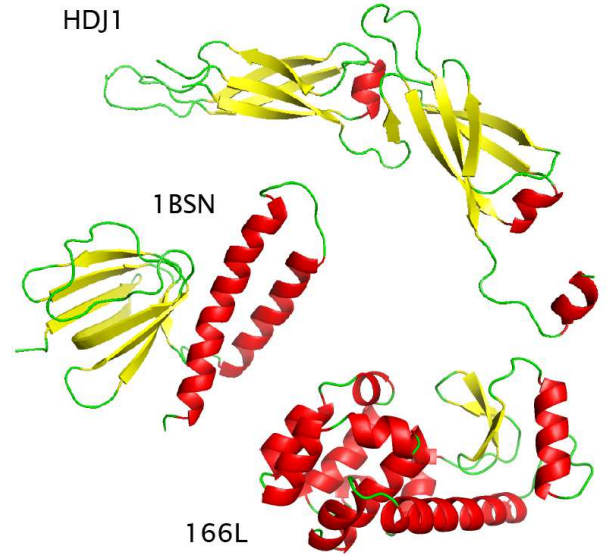} 
\caption{Cartoon representations of three of the analyzed structures, 
along with the corresponding PDB codes.}
\label{f:structures}
\end{figure}
\begin{figure*}[ht!] 
\centering
\subfigure[]{\includegraphics[width=\columnwidth,clip]{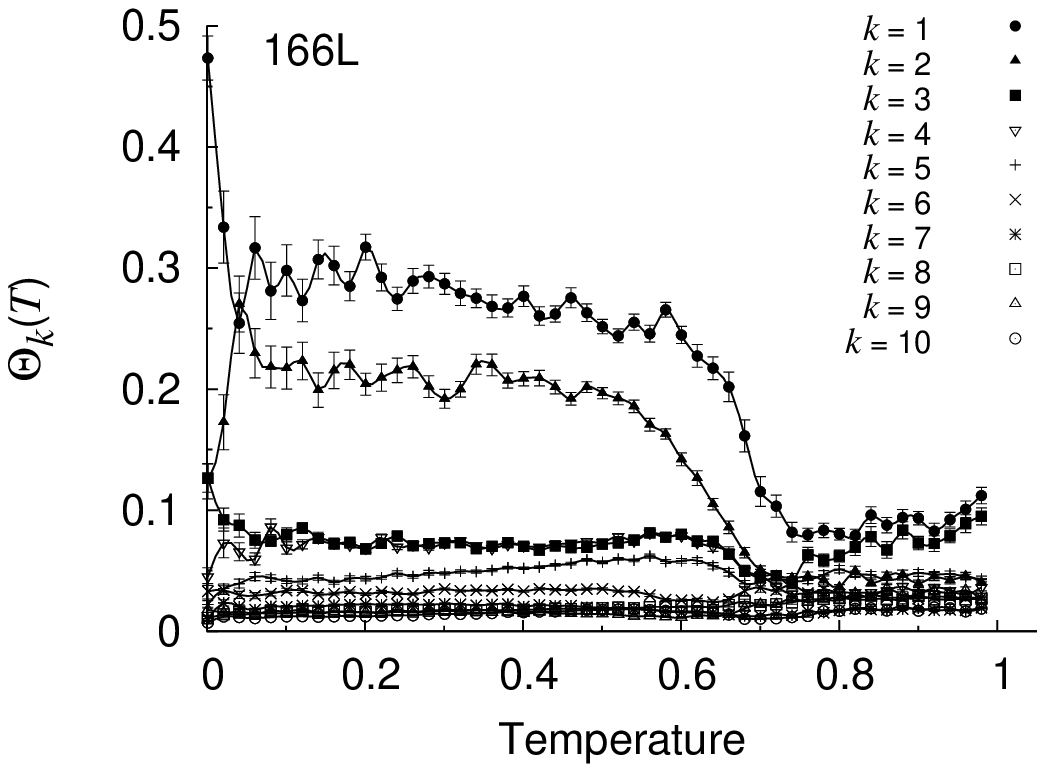}}
\subfigure[]{\includegraphics[width=\columnwidth,clip]{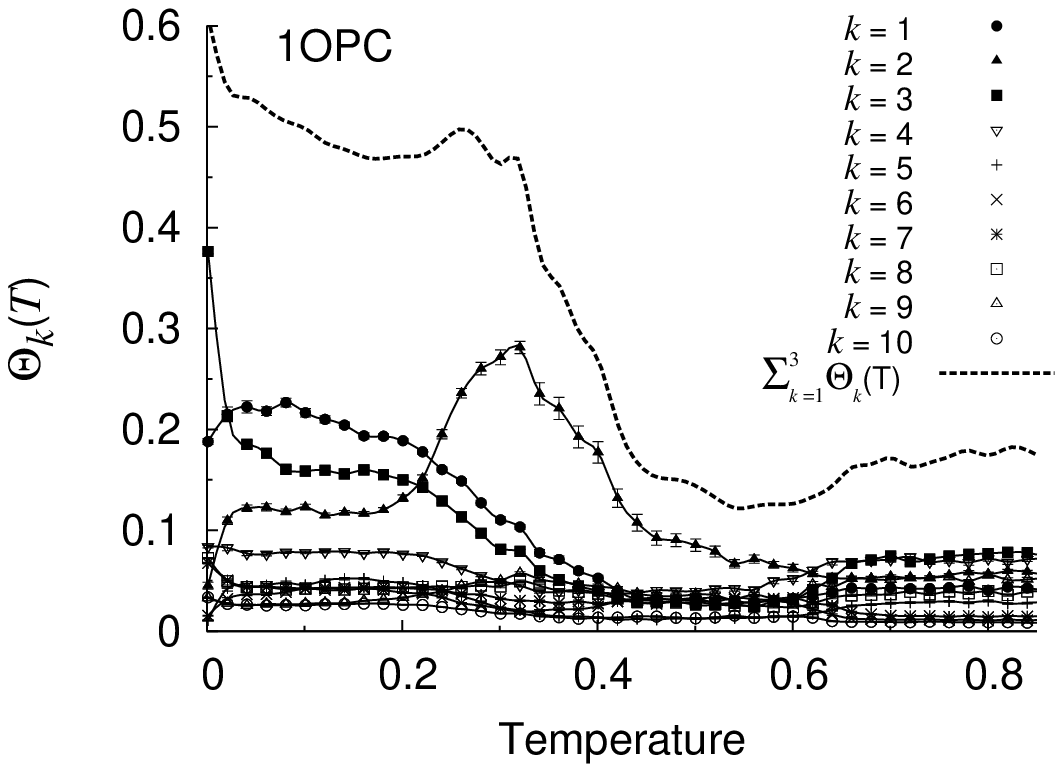}}   
\caption{First ten thermal involvement coefficients of Lysozyme (a), 
$N=162$, $T_{f}=0.75$, PDB code 166L and
of the OMPR DNA-binding domain (b), $N=99$, $T_{f}=0.65$, PDB code 1OPC,
as functions of temperature.}
\label{f:TH3166L}
\end{figure*}

Although NM analysis has been applied with success also to predict large
scale deformations, its appropriate context remains 
the small-amplitude regime and extrapolation to larger fluctuations should 
be rigorously checked to avoid arbitrary interpretations.
In fact, at $T=300$~K,  a mere $10$ \% below typical protein unfolding temperatures ($T_f$), 
displacements well beyond the harmonic 
approximation must be expected.
A possible way to find suitable collective coordinates 
to describe large fluctuations is to use the ``essential modes'',
the eigenvectors of the covariance matrix of the 
displacements from the native state, computed in a molecular
dynamics simulation \cite{Amadei1993}.  
Unfortunately, the convergence of the covariance matrix
toward a stationary solution is very slow~\cite{Balsera1996}.

In this Letter, we establish on a firm basis the validity of the NM
description of protein dynamics and unveil some remarkable subtle features of
this approach as temperature is increased from very low to beyond $T_f$. 


In order to measure the spectral weight of protein
fluctuations as a function of temperature,
we introduce the  thermal involvement coefficients (TICs) $\Theta_{k}(T)$  
\cite{De-Los-Rios2005}. 
They are defined as $\Theta_{k}(T) = \langle  Y_k(T) \rangle_\mathcal{N}$, with
\begin{equation}
\label{e:TICS}
Y_{k}(T) = 
   \frac{\displaystyle \langle  (\overrightarrow{\Delta}(T) \cdot \hat{\psi}_{k})^2 \rangle_{t}}
   {\sum_{k^{\, \prime}}
   \langle(\overrightarrow{\Delta}(T) \cdot \hat{\psi}_{k^{\, \prime}})^2 \rangle_{t}} 
\end{equation}
where $\{\overrightarrow{\Delta}(T) \} = \{ \overrightarrow{r}(T)\} - \{\overrightarrow{R}\}$ 
is the  $3N$-dimensional 
deviation of the protein structure from the native fold configuration
projected onto the NMs,  and averaged both
over time $\langle \dots \rangle_{t}$ and over $\mathcal{N}$  different  
initial conditions drawn from the same equilibrium distribution at temperature $T$, 
$\langle \dots \rangle_\mathcal{N}$. 
We have explicitly verified that the two averages commute, so as to strengthen the statistical 
significance of the sampling and averaging procedures. 
Statistical uncertainties were computed as the 
standard errors on the 
realization averages $\Delta \Theta_{k}(T) =\sqrt{[\langle Y_{k}^2(T) \rangle_\mathcal{N}-
\langle Y_{k}(T) \rangle_\mathcal{N}^2]/\mathcal{N}}$.

Protein trajectories are calculated within the isokinetic scheme~\cite{Morriss1998},
which provides a correct sampling of the 
configuration space~\footnote{Since the 
isokinetic scheme conserves both linear and angular momenta, no special 
alignment procedure is needed before computing conformational changes, if care is taken to set both total momenta to zero at the beginning of each simulation run.}.
Forces are calculated using the coarse-grained G$\bar{\rm{o}}$ model 
introduced in Ref.~\onlinecite{Clementi2000} where each amino-acid is 
replaced by a bead with the average amino-acid mass, whose equilibrium position 
coincides with that of its $\alpha$-carbon.
Successive beads along the chain are connected by stiff harmonic springs,
mimicking the peptide bond and maintaining the chain connectivity.
In line with native-centric schemes,
non-bonded  interactions between non consecutive $\alpha$-carbons 
are modelled with Lennard-Jones 12-10 potentials if the atoms are 
in contact in the native state according to a given interaction cutoff $R_{c}$ 
and with purely repulsive interactions 
otherwise.
The parameters of the non-bonded interactions are 
fixed as $R_c=6.5$~\AA \ and  $R_c=7.5$~\AA \ for $N \leq 100$ and $N>100$ respectively.
The force field is completed by standard harmonic angle-bending  interactions, 
plus the dihedral potential energy
\begin{equation}
\label{eq:dihed}
V_{dh} = 
\sum_{i=3}^{N-2} k_{\phi}^{(1)}[1 - \cos(\Delta \phi_{i})] +
k_{\phi}^{(3)}[1 - \cos3(\Delta \phi_{i})]
\end{equation}
 where $\Delta \phi_{i}$  are the the dihedral angle
 deviations from the native values~\footnote{The dihedral angle at site $i$ is
 defined by the two adjacent planes formed by the four consecutive 
C$_{\alpha}$'s at $i-2$, $i-1$, $i$, $i+1$.}.
\begin{figure*}[t!] 
\centering
\subfigure[]{\includegraphics[width=\columnwidth,clip]{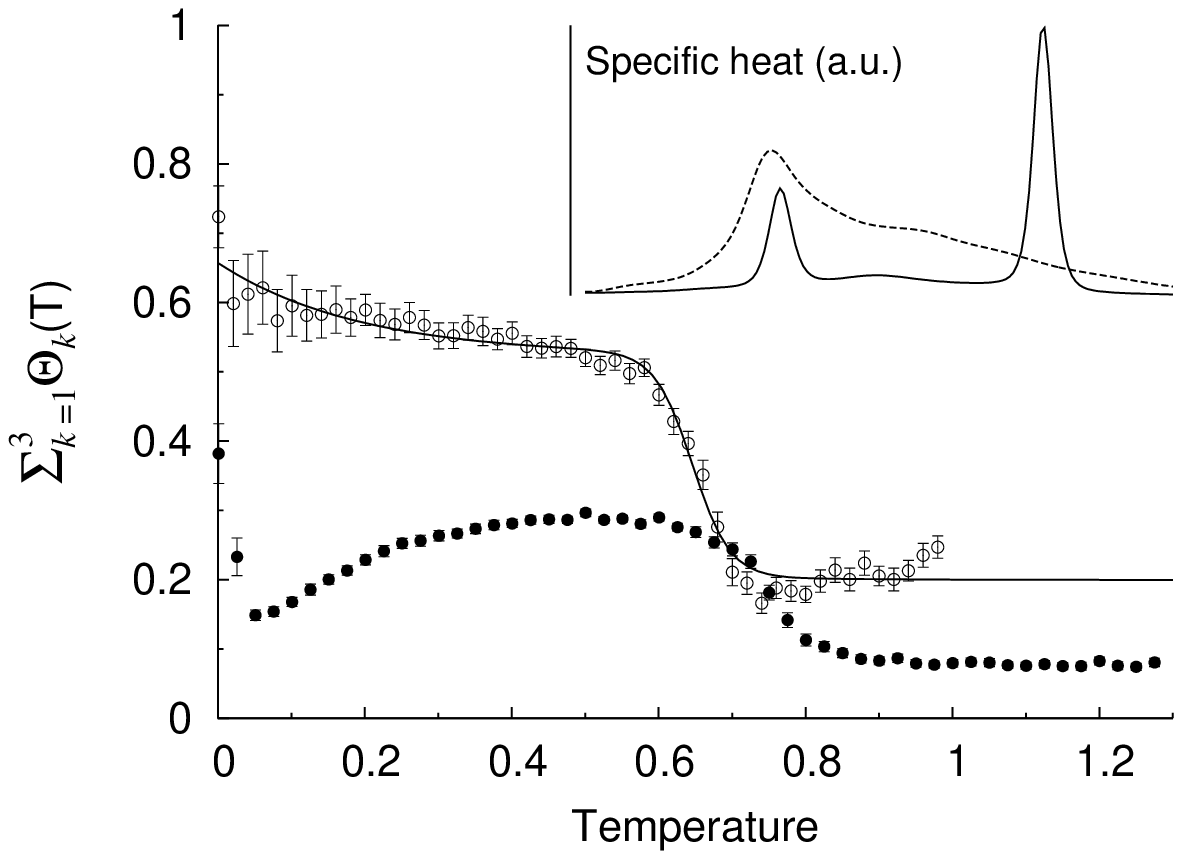}}
\subfigure[]{\includegraphics[width=\columnwidth,clip]{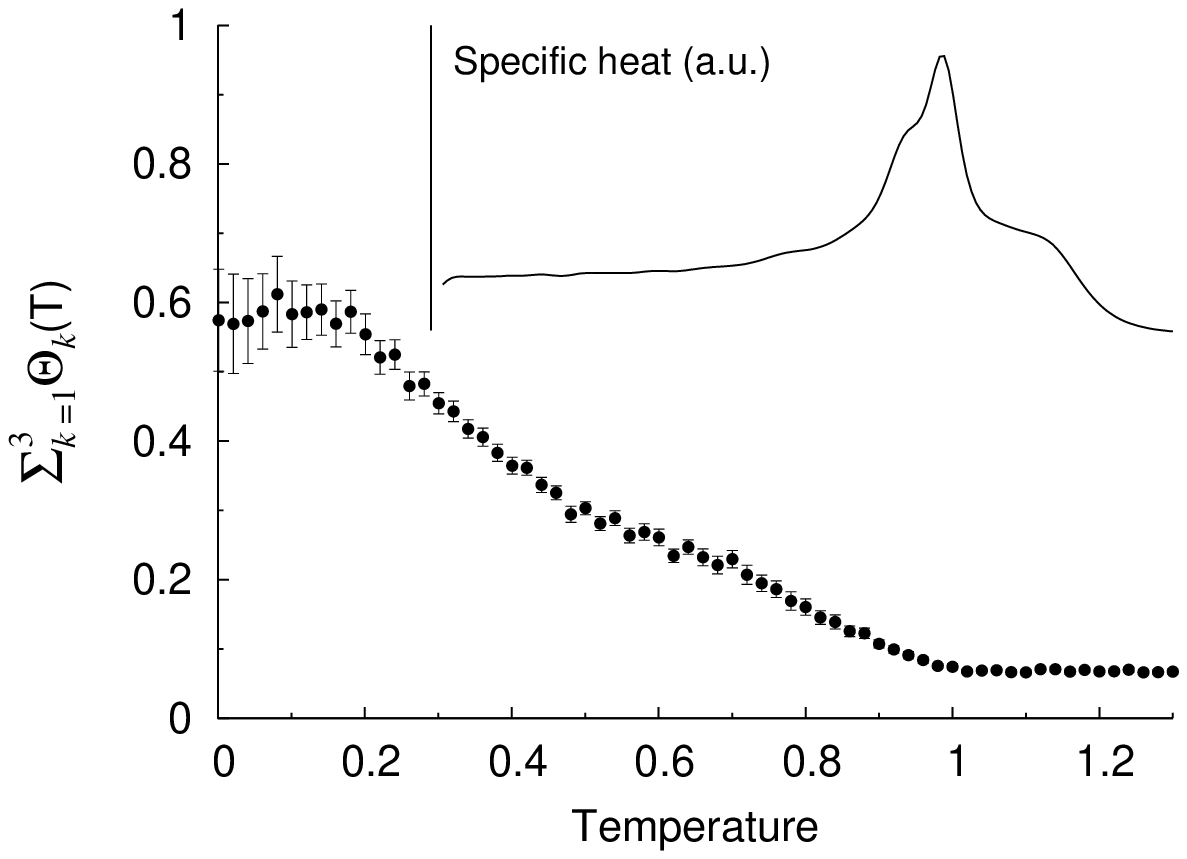}}
\caption{Cumulative involvement coefficient over the first 
three modes for: (a) Lysozyme, $N=162$, PDB code 166L  (empty circles) and a surrogate 
self-avoiding polymer occupying the same volume (filled circles) and (b)
the peptide-binding fragment from Hsp40 (b), $N=171$, PDB code HDJ1.
Insets: specific heat (unfolding) on the same temperature scale.
In panel (a): wild type molecule (dashed line) and the surrogate (solid line).}
\label{f:THHDJ1}
\end{figure*}
Normal modes are computed from the dynamical matrix of the G$\bar{\rm o}$ model, 
according to equation \eqref{eq:harmonic}, after  a conjugate-gradient relaxation of 
the PDB structures, necessary so to obtain the true minimum of the model 
potential~\footnote{Codes available from the authors upon request.}.

In order to locate the unfolding 
temperatures, we have performed thermal unfolding simulations 
coupled to multiple histogram reconstructions~\cite{Ferrenberg1988}
of the specific heat.


We have analyzed a number of protein structures~\footnote{PDB codes: 166L, 1FAS, 1OOI, 
HDJ1, 1BFE, 1BSN, 1CHN, 1FVQ, 1OPC, 1AWO, 1TIT, 1UBI, 1NEB.}, ranging in length from $N=57$
to $N=171$ and heterogeneous for $\alpha$ and $\beta$ content. 
The first notable finding is that a restricted number of low-frequency 
normal modes is always able to capture a sizable fraction of the equilibrium 
fluctuations over a broad
temperature range, from $T=0$ to about 10-20 \% below $T_{f}$. This is clearly 
represented in Fig.\ref{f:TH3166L} showing the behavior of the  first ten TICs of 
Lysozyme and of OMPR DNA-binding domain.
The  TICs of Lysozyme, after a major rearrangement close to $T=0$, 
appear strikingly robust  over the whole range where the protein structure is stable, 
while those of the DNA-binding domain fluctuate wildly. Remarkably, 
in the latter structure, the mode whose TIC is the most important at intermediate 
temperatures gave nearly no contribution to fluctuations 
close to the native minimum. Conversely, the NMs that captured the $T \approx 0$ 
fluctuations appear to swap and rapidly lose spectral weight as temperature increases.
Note that the first TICs of Lysozyme fall off much closer to $T_{f}$ than those of the
DNA-binding domain (see values of $T_{f}$ in the caption). 
These results manifestly warn on the use of single-mode reconstructions of conformational changes using zero-temperature NMs, as these are not
guaranteed to provide stable spectral measures at physiological temperatures.
Indeed the individual low-frequency TICs are observed to fluctuate substantially, 
or even exchange spectral weight among each others, with
respect to zero temperature before stabilizing.  
 
Despite the observed fluctuations of TIC values at low and intermediate 
temperatures, a global measure gathering the contributions from several 
low-frequency NMs
appears to be a more stable indicator of spectral weight 
over the whole thermal span (see Figs.~\ref{f:TH3166L} (b) and~\ref{f:THHDJ1}) for all analyzed 
structures. Note that this is also the case for 
the wildly fluctuating 
TICs of the OMPR DNA-binding domain (see again Fig.~\ref{f:TH3166L} (b)), when the first three of them
are aggregated. Hence, subspace-based, rather than single-mode-based 
reconstructions carry the relevant spectral weight. 
Incidentally, this sheds further light on the observed correlation between
low-frequency subspaces of NMs and essential modes pointed out in 
Ref.~\onlinecite{Micheletti2004}.

As $T_f$ is approached, a  transition of variable width to equipartition is invariably observed. Above $T_f$, TICs are no longer sensitive to structural rearrangements. 
However, the nature of such transitions 
appears rooted in the special structural features of proteins.
More precisely, TICs may relax to equipartition more or less abruptly.
In this regard, crucial issues seem to be 
(i) the degree of structural arrangement at the secondary level and 
(ii) deviations from globularity of the scaffold.

(i) To address the first point, we generated for a few structures self-avoiding polymers 
of the same length as the corresponding proteins and confined in the same enveloping 
ellipsoid as calculated through the corresponding inertial tensor. The necessary excluded 
volume constraints and local persistence length properties were imposed by requiring that 
the three-body radii be never lower than
$2.7$ \AA, as described in Ref.~\onlinecite{Banavar2003}. By doing so, 
the ensuing structures show the same internal
static two- and three-body correlations as the original proteins, while almost lacking the distinctive 
occurrence of $\alpha$ and $\beta$ motifs.
As shown in Fig.~\ref{f:THHDJ1}(a), 
NMs of such surrogate structures do not capture the dynamical fluctuations to the same extent
as in the original proteins. This has been verified for different proteins in our ensemble. 
Hence, we reaffirm the principle that functional motions
that can be captured through 
NM analysis
are rooted in the special 
arrangements of protein folds
~\cite{Tama2001, De-Los-Rios2005, Chennubhotla2005, Tama2006} . 
This is the second important point we wish to stress.

(ii) Not all proteins possess globular folds, some DNA- and peptide- 
binding fragments, for example,  being characterized by less regular shapes, necessary
to adapt to their targets upon binding.  This, in turn, implies a varying degree 
of structure-related 
cooperativity  in the dynamics. Therefore, it is instructive to assess the impact of 
deviations from globularity on the temperature trend of TICs in such cases. 
In Fig.~\ref{f:THHDJ1}(b), we show that the TICs describing 
fluctuations of the human Hsp40 peptide-binding fragment, a 
strongly non-globular protein (see cartoon in Fig. ~\ref{f:structures}), are indeed 
characterized by a weak degree of cooperativity.  In particular, it is clear that 
they no longer represent
a faithful measure of the overall dynamics already well below $T_f$.

\begin{figure}[t!] 
\centering
\includegraphics[width=\columnwidth,clip]{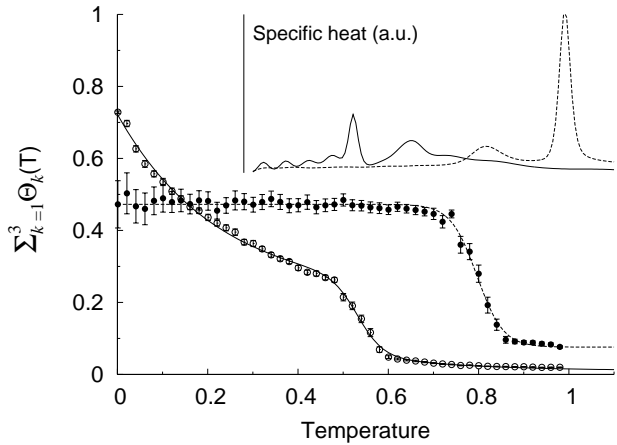} 
\caption{Cumulative involvement coefficient over the first 
three modes for the $\epsilon-$subunit of the F1-ATPase ($N=138$, PDB code 1BSN) computed with the full 
potential (filled circles) and without dihedral forces (empty circles). Lines are just
guides to the eye.  Inset: specific heat (unfolding)
for the full model (dashed line) and without dihedral terms (solid line) on the same temperature scale.}
\label{f:TH31BSN}
\end{figure}


Finally, to highlight  the role of the energy landscape features on the temperature dependence of TICs, we have analyzed
F1-ATPase, a globular protein with a high content  of  of $\alpha$ and $\beta$ motifs. 
Fig.~\ref{f:TH31BSN} shows the TIC thermal plot computed both with the full 
potential and by switching off the dihedral terms~\eqref{eq:dihed}, which are
essential in order to reproduce the cooperativity of protein folding 
thermodynamics within a G$\bar{\rm{o}}$ scheme~\cite{Knott2004}.
Since it is also known that less globular proteins are less cooperative, 
we expect that, due to the lack of the dihedral term, TICs will also be affected, 
showing a behavior similar to what observed for
non-globular proteins such as HDJ1. Indeed, Fig.~\ref{f:TH31BSN} documents the validity 
of our inference. Note that the strong cooperativity and thermal resilience of TICs are both 
profoundly eroded starting from relatively small temperatures.
This completes  the series of our findings.

In this Letter we have shown
that sizable values and thermal 
resilience of TICs are deeply rooted in the 
presence of secondary motifs and in the degree of their packing within protein scaffolds.
The fluctuations of more irregular, less compact and less cooperative proteins are
manifestly harder to capture through normal modes. 
Although our results confirm the validity of NM analysis of protein functional motions, 
we have shown that subspaces, typically spanned by the first few low-frequency modes, 
rather than single normal modes, ought to be 
considered as identifying the relevant functional motions. 
This conclusion stems in a clear fashion from the 
observation that the {\it spectral capacity}  of individual 
modes, as measured through TICs, often differs substantially at physiological 
temperatures with respect to the traditional 
static involvement coefficients at zero-temperature.
Moreover, we have brought to the fore a crucial {\em liaison}  between the
meaningfulness of a NM-based description of functional fluctuations at physiological 
temperatures and the degree of cooperativity of  proteins.

The authors wish to thank Bingdong Sha for making the HDJ1 structure available
prior to publication.


\end{document}